# HIGH-ENERGY EMISSION FROM PRESUPERNOVAE

Volodymyr Kryvdyk[1]

[1] *Depart. of Astronomy, Kyiv University, av. Glushkova 6, Kyiv 03022, Ukraine*

## ABSTRACT

The non-thermal emission in the magnetospheres of presupernova collapsing stars with initial dipole magnetic fields and a certain initial energy distribution of the charged particles in a magnetosphere is considered. The analysis of particle dynamics and its emission in the stellar magnetosphere under collapse show that the collapsing stars can be the powerful sources of a non-thermal radiation produced by the interaction of charged particles with the magnetic field. The radiation flux grows with decreasing stellar radius. This flux can be observed in the form of radiation burst with duration equal to the stellar collapse time. The radiation flux depends on the distance to the stars, the particle spectrum and magnetic field in the magnetosphere. The radiation flux is calculated for various collapsing stars with initial dipole magnetic fields and an initial power series, relativistic Maxwell, and Boltzmann particle energy distribution in the magnetosphere. This radiation can be observed by means of modern astronomical instruments.

## INTRODUCTION

There are three ways to observe the stars on the stage of gravitational collapse, namely to detect 1) neutrinos, 2) the gravitational waves, and 3) electromagnetic radiation, arisen by collapse.

The first neutrino signal from supernova stars was detected in 1987 from SN 1987 A (Bionta, et al., 1987, Hirata , et al., 1987). It is the unique observation of a star on the stage of its explosion.

The stars must emit electromagnetic and gravitational waves under the collapse. But these electromagnetic waves are very low frequencies (Cunningam , et al. , 1978, 1979, 1980, Henriksen, et al.,1979, Moncrief ,1980) therefore their can not observed near Earth. Still, these waves are not detected, and maybe this is the main problem in the theory of stellar collapse.

In this paper the one with method for the observation of the stellar collapse is proposed using the radiation arisen in magnetospheres of collapsing stars. This radiation will be generated when the star magnetosphere compress under the collapse and its magnetic field considerably increases. Thus a cyclic electric field is produced, involving of acceleration charged particles that generate radiation when moving in the magnetic field. The frequencies of this radiation are very high (from gamma-rays to radio waves) therefore they can be observed near Earth

## COLLAPSAR MAGNETOSPHERE

We considered the collapse of stars with an initial dipolar magnetic field and three typical particles distributions in the magnetosphere: 1) power-series $N_1(E) = K_p E^{-\gamma}$; 2) relativistic Maxwell $N_2(E) = K_M E^2 \exp(-E/kT)$; and 3) Boltzmann distributions $N_3(E) = K_B \exp(-E/kT)$

The external electromagnetic field of the collapsing stars will change as (Ginzburg and Ozernoy, 1964)

$$\begin{aligned}
B_r &= 2r^{-3}\mu(t)\cos\theta \ , \\
B_\theta &= r^{-3}\mu(t)\sin\theta \ , \\
B_\varphi &= 0 \\
E_\varphi &= -c^{-1}r^{-2}\frac{\partial \mu}{\partial t}\sin\theta \ ,
\end{aligned} \qquad (1)$$

$$E_r = E_\theta = 0,$$

where μ(t) is the variable magnetic momentum of the collapsar.

The field structure and particles dynamics in the magnetosphere are influenced by three factors: particles pressure, collisions, and star rotation. As follow from a detailed analysis (Kryvdyk, 1999) these effects can be neglected during the collapse.

In order to investigate the particle dynamics during the collapse we use the method of adiabatic invariant. In this case the particles energy will change as results of the two mechanisms: a betatron acceleration in the variable magnetic field, 2) bremsstrahlung energy losses in this field.

For the betatron acceleration in drift approximation we can write (Bakhareva and Tverskoj, 1981)

$$(\frac{dE}{dt})_a = -(1/3) pv \, div \vec{u} \qquad (2)$$

where $p$ and $v$ is particle impulse and particle velocity, $u = cB^{-2}[\vec{E}\vec{B}]$ is the drift velocity.

For the dipolar magnetic field (1)

$$(\frac{dE}{dt})_a = (4/3)\mu^{-1}(\partial \mu / \partial t) pv f(\theta)$$

where $f(\theta) = (3\cos^4\theta - 1)(1 + 3\cos^2\theta)^{-2}$.

The magnetic moment μ(t) of collapsar change in a results of the change their radius R(t) under the influence of gravitational field according to the law of free fall

$$\frac{dR}{dt} = (2GM(R_* - 1)/RR_*)^{1/2} \quad .$$

Here $R_* = R_0/R$.

Passing to the new variable $R = R(t)$ we can write for the betatron acceleration

$$(\frac{dE}{dt})_a = \frac{4}{3} k_1 (2GM/R^3)^{1/2} [(R_0 - R)/R_0]^{1/2} f(\theta) E$$

Here $k_1 = 2$ and $k_1 = 1$ for relativistic and non-relativistic particles respectively.

For the bremsstrahlung energy losses (Ginzburg and Syrovatskij, 1964)

$$(\frac{dE}{dt})_s = (e^4/6m^4c^7)(B_0 R_0)^2 g(\theta, \alpha) R^2 E^2 r^{-6},$$

where α is the angle between the magnetic field and the particle velocity.

Here $g(\theta, \alpha) = (1 + 3\cos^2\theta)\sin^2\alpha$.

The resulting rate of particle energy change in the magnetosphere is

$$(\frac{dE}{dt}) = (\frac{dE}{dt})_a + (\frac{dE}{dt})_s = \tfrac{4}{3} k_1 (2GM(R_* - 1)/R_* R^3)^{1/2} f(\theta) E - (e^4/6m^4c^7)(\Phi_0)^2 g(\theta, \alpha) R^2 E^2 r^{-6} \qquad (3)$$

Particle dynamics can be investigated using the equation of transitions particle in the regular magnetic fields (Ginzburg and Syrovatskij, 1964)

$$\frac{\partial N}{\partial t} + \frac{\partial}{\partial E}(N \frac{dE}{dt}) = 0$$

For the new variable $R = R(t)$ this equation becomes

$$\frac{\partial N}{\partial R} = f_1(E,R)\frac{\partial N}{\partial E} + f_2(E,R)N = 0, \qquad (4)$$

Here

$$f_1(E,R) = ER^{-1}\{A_1 - A_2 R^3 [R/(R_0 - R)]^{1/2} E\};$$
$$f_2(E,R) = R^{-1}\{A_1 - 2R^3 [R/(R_0 - R)]^{1/2} E\},$$
$$A_1 = 4k_1 f(\theta)/3;$$
$$A_2 = (e^4/6m^4c^7)(B_0 R_0^2)^2 (R_0/2GM)^{1/2} g(\theta,\alpha)$$

Eq. (4) can not be solved in the general case and so two special cases are considered: (i) when energy losses do not influence the particle spectrum spectrum in the magnetosphere and (ii) when the energy losses determine the particle spectrum .
The solution of Eq. (4) in these two cases is given by

$$N^i_1(E,R) = K_p E^{-\gamma} R_*^{-\beta_1};$$
$$N^i_2(E,R) = K_M E^2 R_*^{-\beta_2} \exp(-E/kT) \qquad (5)$$
$$N^i_3(E,R) = K_B R_*^{-\beta_3} \exp(-E/kT) ;$$

$$N^{ii}_1(E,R) = K_p \exp(-\gamma(1-\gamma_1));$$
$$N^{ii}_2(E,R) = K_M E^2 \exp(-(1-\gamma_1)E/kT); \qquad (6)$$
$$N^{ii}_3(E,R) = K_B \exp(-(1-\gamma_1)E/kT) .$$

Here $\gamma_1 = A_2 F(R,R_*)r^{-6}E$; $\beta_P = A_1(\gamma_0 - 1)$; $\beta_M = A_1(E/kT \ln E - 3)$; $\beta_B = A_1(E/kT \ln E - 1)$

Eqs. (5) determine the particle spectrum in the magnetosphere and its evolution during the initial stage of the collapse when the energy losses can be neglected. We will consider this case in this paper. Eqs. (6) determine the particles spectrum on the final stage of the collapse, when the magnetic field attains an extreme value and the energy losses influence the particle spectrum considerably. This case we will consider in a later paper.

**NON-THERMAL EMISSION FROM COLLAPSAR**

The ratio between the radiation flux from collapsar with radius $R$ and its initial radiation flux (by $R = R_0$) for the power-series has been obtained previously in the paper ( Kryvdyk,1999)

$$I_{\nu P}/I_{\nu P0} = (\nu/\nu_0)^{(1-\gamma)/2} R_*^{\gamma-2} \int_0^{\pi/2}\int_0^{\infty} R_*^{A_1(\gamma-2)} \sin\theta d\theta dE, \qquad (7)$$

Similarly for relativistic Maxwell and Boltzmann distributions we can write down

$$I_{\nu M}/I_{\nu M0} = (\nu/\nu_0) R_*^{-3} (1/kT) \int_0^{\pi/2}\int_0^{\infty} R_*^{-\beta_M} \exp(E/kT) \sin\theta d\theta dE, \qquad (8)$$

$$I_{\nu B}/I_{\nu B0} = (\nu/\nu_0) R_*^{-3} (kT) \int_0^{\pi/2}\int_0^{\infty} R_*^{-\beta_B} E^{-2} \exp(E/kT) \sin\theta d\theta dE, \qquad (9)$$

Using Eq. (7)-(9) the radiation flux from the collapsar magnetospheres with variable dipole magnetic fields can be calculated. The ratio between the radiation flux from a collapsars with radius $R$ and their initial flux by $\nu/\nu_0 = 1$ are:

$$1 \leq I_{\nu P}/I_{\nu P0} \leq 1.34 \cdot 10^{10} \quad \text{for } 2.4 \leq \gamma \leq 3.4, \quad 10 \leq R_* \leq 1000,$$
$$1 \leq I_{\nu B}/I_{\nu B0} \leq 4.86 \cdot 10^5 \quad \text{for } 1\,eV \leq kT \leq 9eV, \quad 145 \leq R_* \leq 850,$$
$$1 \leq I_{\nu M}/I_{\nu M0} \leq 2.23 \cdot 10^{11} \quad \text{for } 1\,eV \leq kT \leq 9eV, \quad 145 \leq R_* \leq 850.$$

These values obtained by the numerical integration of the equations (7)-(9) for $2\ eV \leq E \leq 10^9\ eV$ and the different $R_*, kT, \gamma$.

**CONCLUSIONS**

The magnetic fields of presupernova stars increase during the collapse. These variable magnetic fields accelerate the charged particles in the magnetospheres of collapsing stars. These particles emit the electromagnetic waves in the wide frequency band, from gamma rays to radio waves. The radiation flux grows during collapse and to reaches a maximum on the final stage of collapse. This radiation can be observed as bursts in the all-frequency band, from radio to gamma ray. The burst duration completes with the collapse time. The radiation flux from collapsar on the final stage of collapse exceeds the initial flux a million times. Thus the collapsar can be the powerful sources of the non-thermal radiation when before a supernova flares, the star compresses and emits bursts. Where can we seek these bursts? First of all in powerful gamma bursts and X bursts which are not periodical..

What problems can arise in the program for the astrophysical observation of bursts from presupernova collapsing stars? Above all now the theories of the stellar evolution do not enable us to locate the presupernova collapsar with enough accuracy. We can indicate only on the stars types that can collapse to supernova, but we can not point to its location. This fact is principal problem in the observational astrophysics program for the search of collapsing presupernova stars. A next problem is how to choose the impulse from presupernova collapsar from the great number of the bursts with unknown origin.